# Analytical simulations of the resonant transmission of electrons in a closed nanocircuit for terahertz applications where a tunneling junction is shunted by a metallic nanowire.

Mark J. Hagmann

*Department of Electrical and Computer Engineering, University of Utah, 50 S. Central Campus Dr #2110, Salt Lake City, Utah 84112, USA*

*Corresponding author: hagmann@ece.utah.edu*

**ABSTRACT**

Earlier, in the CINT program at Los Alamos National Laboratory, we focused ultrafast mode-locked lasers on the tip-sample junction of a scanning tunneling microscope to generate currents at hundreds of harmonics of the laser pulse repetition frequency. Each harmonic has a signal-to-noise ratio of 20 dB with a 10-dB linewidth of only 3 Hz. Now we model closed quantum nanocircuits with rectangular, triangular, or delta-function barrier, shunted by a beryllium filament for quasi-coherent electron transport over mean-free paths as great as 68 nm. The time-independent Schrödinger equation is solved with the boundary conditions that the wavefunction and its derivative are continuous at both connections. These four boundary conditions are used to form a four-by-four complex matrix equation with only zeros in the right-hand column vector which is required to have a non-trivial solution with each of the closed nanocircuits. Each model has four parameters: (1) the barrier length, (2) the height and shape of the barrier, (3) the length of the pre-barrier, and (4) the electron energy. Any three of these may be specified and then the fourth is varied to bring the determinant to zero to find the solutions on lines or surfaces in the space defined by the four parameters. First, we use a simplistic model having a rectangular barrier. The second model has a triangular barrier as a first approximation to field emission, and we are considering applying this approach for a self-contained nanoscale extension of our earlier effort to generate the harmonics at Los Alamos. The third model has a delta-function barrier, and the fourth model is an extension of the first one where the width of the rectangular barrier is varied inversely with its height.

## I. INTRODUCTION

We derived analytical solutions of the time-independent Schrödinger equation for three types of closed quantum nanocircuits. Each nanocircuit has a tunneling junction connected across a pre-barrier region that is at zero potential. Each model shows the tunneling junction and the pre-barrier on a straight line with one connection made at their point of contact. The second connection is shown by two ground symbols at the outer ends of the model.

The first model has a rectangular barrier and the second model has a triangular barrier to approximate the effects of field emission. The rectangular-barrier model could also be used to represent the injection of free electrons into the barrier. We consider the second model with a triangular barrier to be more realistic and plan to make and testing prototypes. The third model has a delta-function barrier. The fourth model has a pre-barrier region with a rectangular barrier where the height varies inversely with the length to show the transition from the first model to the third.

The pre-barrier region could be a filament of a metal such as beryllium for which electrons have a mean-free path as great as 68 nm for quasi-coherent transport. The matrix equation for each model consists of four linear equations with a set of four parameters with four coefficients as the unknowns. When three coefficients are specified the fourth one is determined.

## II. CLOSED-CIRCUIT MODEL WITH A RECTANGULAR BARRIER

Figure 1 is a diagram showing the potential for a one-dimensional model consisting of a rectangular barrier with a pre-barrier region. Note that the parameter a must be less than zero.

Models with rectangular barriers have been used when introducing quantum tunneling [1] but these models are heuristic and it appears that they do not relate to any experiments or applications. Note that the electric field is infinite at both vertical edges of a rectangular barrier. However, for the model in Fig. 1, the two lower ends of the barrier are connected at a single point, shown by the two ground symbols, to complete the circuit. Thus, this model represents a circular ring having an opening in Region II with no electric field so that electrons are injected and collected by a separate system.

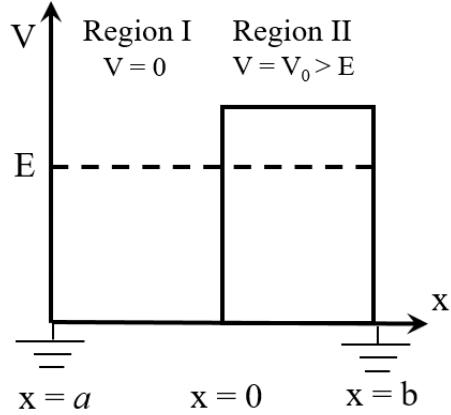

Fig. 1. Potential energy for $a$ closed circuit with a rectangular barrier.

We write the time-independent Schrödinger equation as Eq. 1, where E is the energy of the particle and V(x) is the potential energy.

$$\frac{d^2\psi}{dx^2} + \frac{2m}{\hbar^2}[E-V(x)]\psi = 0 \qquad (1)$$

Solving the Schrödinger equation, we obtain Equations 2 and 3 for the wavefunctions in regions I and II where the parameters k and β are defined in Eqns. 4 and 5. Taking the derivatives in respect to x gives Eqns. 6 and 7.

$$\psi_I = A\cos(kx) + B\sin(kx) \qquad (2)$$

$$\psi_{II} = Ce^{\beta x} + De^{-\beta x} \qquad (3)$$

$$k \equiv \frac{\sqrt{2mE}}{\hbar} \qquad (4)$$

$$\beta \equiv \frac{\sqrt{2m(V_0 - E)}}{\hbar} \qquad (5)$$

$$\frac{d\psi_I}{ds} = -kA\sin(kx) + kB\cos(kx) \qquad (6)$$

$$\frac{d\psi_{II}}{dx} = \beta Ce^{\beta x} - \beta De^{-\beta x} \qquad (7)$$

Applying the boundary conditions so that the wavefunction and its derivative are continuous for x equal to zero gives Eqns. 8 and 9.

$$A - C - D = 0 \qquad (8)$$

$$kB - \beta C + \beta D = 0 \qquad (9)$$

Requiring continuity of the wave function and its derivative at the common boundary where x equals $a$ in region I to that where x equals b in region II gives Eqns. 10 and 11.

$$A\cos(ka) + B\sin(ka) - Ce^{\beta b} - De^{-\beta b} = 0 \qquad (10)$$

$$kA\sin(ka) - kB\cos(ka) + \beta Ce^{\beta b} - \beta De^{-\beta b} = 0 \qquad (11)$$

The system of Eqs, 8, 9, 10, and 11 in the four coefficients A, B, C, and D may be written in matrix form as shown in Eq. 12.

$$\begin{bmatrix} +1 & 0 & -1 & -1 \\ 0 & +k & -\beta & +\beta \\ +\cos(ka) & +\sin(ka) & -e^{\beta b} & -e^{-\beta b} \\ +k\sin(ka) & -k\cos(ka) & +\beta e^{\beta b} & -\beta e^{-\beta b} \end{bmatrix} \begin{bmatrix} A \\ B \\ C \\ D \end{bmatrix} = \begin{bmatrix} 0 \\ 0 \\ 0 \\ 0 \end{bmatrix} \qquad (12)$$

However, this is a homogeneous system of equations so to have a non-trivial solution for the four coefficients the determinant must be zero, as shown in Eq. 13.

$$\begin{vmatrix} +1 & 0 & -1 & -1 \\ 0 & +k & -\beta & +\beta \\ +\cos(ka) & +\sin(ka) & -e^{\beta b} & -e^{-\beta b} \\ +k\sin(ka) & -k\cos(ka) & +\beta e^{\beta b} & -\beta e^{-\beta b} \end{vmatrix} = 0 \qquad (13)$$

In general, expanding a determinant with four rows and four columns will give 24 terms. However, for the determinant in Eq. 13, ten terms are zero to reduce this to 14 terms. Also setting the matrix elements that are plus or minus 1 in these 14 terms gives the simpler expression in Eq. 14 for the determinant.

$$Det = -M_{23}M_{32}M_{41} + M_{24}M_{32}M_{41} + M_{22}M_{33}M_{41} - M_{22}M_{34}M_{41} + M_{23}M_{31}M_{42}$$
$$- M_{24}M_{31}M_{42} - M_{24}M_{33}M_{42} + M_{23}M_{34}M_{42} - M_{22}M_{31}M_{43} + M_{24}M_{32}M_{43}$$
$$- M_{22}M_{34}M_{43} + M_{22}M_{31}M_{44} - M_{23}M_{32}M_{44} + M_{22}M_{33}M_{44} \qquad (14)$$

Entering the expressions for each of the remaining terms in Eq. 14, introducing hyperbolic functions, and simplifying gives Eq. 15.

$$Det = 4k\beta + 2(\beta^2 - k^2)\sinh(\beta b)\sin(ka) - 4k\beta\cosh(\beta b)\cos(ka) \qquad (15)$$

Defining the angles $\Theta$ and $\Phi$ in Eqs. 16 and 17, we obtain Eq. 18 for the determinant divided by non-zero quantity of $2k\beta$. Thus, the right-hand side of Eq. 18 must be zero for a nontrivial solution. Note that because $a$ is negative $\Theta$ is non-positive, whereas $\Phi$ is non-negative.

$$\Theta \equiv ka \qquad (16)$$

$$\Phi \equiv \beta b \qquad (17)$$

$$\frac{Det}{2k\beta} = \left(\frac{\beta}{k} - \frac{k}{\beta}\right)\sinh(\Phi)\sin(\Theta) + 2[1 - \cosh(\Phi)\cos(\Theta)] \qquad (18)$$

We use the following algorithm with Eq. 18 to determine the pre-barrier length as a function of the energy and the length of the rectangular barrier when the charge and mass of the electron are specified. Simulations for this model are presented in Table I of section IX. Note that, for each value for the particle energy the pre-barrier length is proportional to the barrier length. Also, at each value for the barrier length the pre-barrier length is proportional to the particle energy.

(1) Specify the particle energy E.
(2) Specify the potential energy $V_0$.
(3) Specify the barrier length as the positive number b.
(4) Specify the angle Θ, as a negative number of radians.
(5) Determine the sine and cosine of Θ.
(6) Use E in Eq. 4 to determine k.
(7) Use of E and $V_0$ in Eq. 5 to determine β.
(8) Use b and β in Eq. 17 to determine Φ.
(9) Use k, β, Θ, and Φ in Eq. 18 to obtain the determinant.
(10) Return to step 4 to enter a new value of Θ and continue this process to bring the determinant to zero thus obtaining the complete solution for the specified values of E, $V_0$, and b. at this point $a$ may be determined by dividing Θ by k.

## III. COEFFICIENTS FOR THE RECTANGULAR BARRIER WAVEFUNCTION.

We normalize the system of four equations by defining A to be unity in Eq. 19 and modify the set of four equations by dividing Eqns. 9 and 11 by k to obtain Eqns. 20, 21, 22, and 23.

$$A \equiv 1 \tag{19}$$

$$C + D = 1 \tag{20}$$

$$B - \frac{\beta}{k}C + \frac{\beta}{k}D = 0 \tag{21}$$

$$\cos(ka) + B\sin(ka) - Ce^{\beta b} - De^{-\beta b} = 0 \tag{22}$$

$$\sin(ka) - B\cos(ka) + \frac{\beta}{k}Ce^{\beta b} - \frac{\beta}{k}De^{-\beta b} = 0 \tag{23}$$

Combining Eqs. 20 and 21 we obtain the following solutions for C and D:

$$C = \frac{1}{2}\left(1 + \frac{k}{\beta}\right) \tag{24}$$

$$D = \frac{1}{2}\left(1 - \frac{k}{\beta}\right) \tag{25}$$

Substituting Eqs. 24 and 25 into Eq. 23

$$B = \frac{\beta}{2k\cos(ka)}\left(1 + \frac{k}{\beta}\right)e^{\beta b} - \frac{\beta}{2k\cos(ka)}\left(1 - \frac{k}{\beta}\right)e^{-\beta b} + \tan(ka) \tag{26}$$

Thus, the coefficients B, C, and D may be determined with Eqns. 24, 25, and 26. These equations may be used to determine the coefficients at the end of the algorithm which is defined at the end of the previous section.

## IV. CLOSED-CIRCUIT MODEL WITH A TRIANGULAR BARRIER

Landau and Lifshitz [2] presented an exact solution of the Schrödinger equation for the one-dimensional motion of a particle in a homogeneous external field in terms of Airy functions that was first published in 1965. Now we present a solution for a triangular barrier having finite linear extent so it is necessary to include both Airy functions with their derivatives in the solution. This one-dimensional closed-circuit model is shown in Fig. 2. As in Fig. 1, the two ground symbols represent a common connection with zero length to complete the closed circuit. Note that the potential $V_0$ has units of energy and the coordinate labeled "a" has a negative value. The left edge of the barrier represents an ideal voltage source.

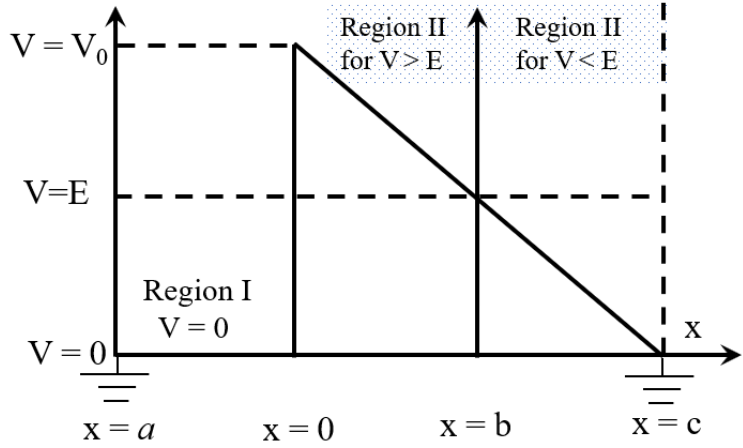

Fig. 2. Potential energy in regions I and II with a triangular barrier

To keep the derivations for the first and second models separate we use Eq. 27 for the wavefunction in Region I with the definition of k in Eq. 4. The coefficients A and B are constants to be determined by the boundary conditions.

$$\psi_I = A\cos(kx) + B\sin(kx) \qquad (27)$$

In Region II Airy functions [1] are required to solve the time-independent Schrödinger equation in Eq. 28 to determine the wave function where E and V have units of electron volts and e is the magnitude of the charge on an electron.

$$\frac{d^2\psi_{II}}{dx^2} + \frac{2me}{\hbar^2}[E - V(x)]\psi_{II} = 0 \qquad (28)$$

The potential energy within the barrier is given by Eq. 29 as a linear interpolant. Thus, the x coordinate b at which the potential energy is equal to the particle energy is given by Eq. 30.

$$V(x) = \left(1 - \frac{x}{c}\right)V_0 \qquad (29)$$

$$b = \left(1 - \frac{E}{V_0}\right)c \qquad (30)$$

Substituting the expression for the potential energy from Eq. 29 into Eq. 28 gives Eq. 31.

$$\frac{d^2\psi_{II}}{dx^2} + \frac{2me}{\hbar^2}\left[(E - V_0) + V_0\frac{x}{c}\right]\psi_{II} = 0 \qquad (31)$$

Next, we make a change of the parameters which is similar to that made in the half-space solution by Valée and Soares [1] to obtain Eq. 32 where the argument ξ is defined in Eq. 33. The coefficients C and D are constants to be determined by the boundary conditions.

$$\psi_{II}(x) = CAi(\xi) + DBi(\xi) \tag{32}$$

$$\xi \equiv \left(\frac{2mV_0 e}{\hbar^2 c}\right)^{\frac{1}{3}} \left(\frac{E}{V_0} - 1\right) c + \left(\frac{2mV_0 e}{\hbar^2 c}\right)^{\frac{1}{3}} x \tag{33}$$

We separate the argument ξ into two parts as shown in Eq. 34 where γ and K are defined in Eqns. 35 and 36. Thus, the wavefunction and its derivative in Region II are given by Eqns. 37 and 38. Note that γ is greater than zero, and K is negative for E less than $V_0$ in quantum tunneling.

$$\xi = K + \gamma x \tag{34}$$

$$\gamma \equiv \left(\frac{2mV_0 e}{\hbar^2 c}\right)^{\frac{1}{3}} \tag{35}$$

$$K \equiv -\left(1 - \frac{E}{V_0}\right)\gamma c \tag{36}$$

$$\psi_{II}(x) = CAi(K + \gamma x) + DBi(K + \gamma x) \tag{37}$$

$$\psi_{II}'(x) = \gamma CAi'(K + \gamma x) + \gamma DBi'(K + \gamma x) \tag{38}$$

From Eq 27, in Region I the derivative of the wavefunction is given by Eq. 39.

$$\psi_I'(x) = -kA\sin(kx) + kB\cos(kx) \tag{39}$$

Thus, the wavefunctions and their derivatives just inside the two ends of Region I are given by Eqns. 40 to 43.

$$\psi_I(0^-) = A \tag{40}$$

$$\psi_I'(0^-) = kB \tag{41}$$

$$\psi_I(a^+) = \cos(ka)A + \sin(ka)B \tag{42}$$

$$\psi_I'(a^+) = -k\sin(ka)A + k\cos(ka)B \tag{43}$$

The wavefunctions and their derivatives just inside the two ends of Region II are given by Eqns. 44 to 47.

$$\psi_{II}(0^+) = Ai(K)C + Bi(K)D \tag{44}$$

$$\psi_{II}'(0^+) = \gamma Ai'(K)C + \gamma Bi'(K)D \tag{45}$$

$$\psi_{II}(c^-) = Ai(K + \gamma c)C + Bi(K + \gamma c)D \tag{46}$$

$$\psi_{II}'(c^-) = \gamma Ai'(K + \gamma c)C + \gamma Bi'(K + \gamma c)D \tag{47}$$

Next pairs from the system of Eqns. 40 to 47 are used to form the following system of 4 simultaneous equations in the 4 unknown coefficients as required for continuity of the wavefunctions and their derivatives.

From Eqns. 40 and 44, for $\psi_I(0^-)$ equal to $\psi_{II}(0^+)$:

$$A = Ai(K)C + Bi(K)D \tag{48}$$

From Eqns. 41 and 45, for $\psi_I'(0^-)$ equal to $\psi_{II}'(0^+)$:

$$kB = \gamma Ai'(K)C + \gamma Bi'(K)D \quad (49)$$

From Eqns. 42 and 46, for $\psi_I$ ($a^+$) equal to $\psi_{II}$ ($c^-$):

$$\cos(ka)A + \sin(ka)B = Ai(K+\gamma c)C + Bi(K+\gamma c)D \quad (50)$$

From Eqns. 43 and 47, for $\psi_I'$ ($a^+$) equal to $\psi_{II}'$($c^-$):

$$-k\sin(ka)A + k\cos(ka)B = \gamma Ai'(K+\gamma c)C + \gamma Bi'(K+\gamma c)D \quad (51)$$

Next, we rearrange these four equations as expressions with zero on the right-hand side and divide those containing $\gamma$ by k so that they are dimensionless, where the parameter R is defined in Eq. 52.

$$R \equiv \frac{\gamma}{k} \quad (52)$$

$$A - Ai(K)C - Bi(K)D = 0 \quad (53)$$

$$B - RAi'(K)C - RBi'(K)D = 0 \quad (54)$$

$$\cos(ka)A + \sin(ka)B - Ai(K+\gamma c)C - Bi(K+\gamma c)D = 0 \quad (55)$$

$$\sin(ka)A - \cos(ka)B + RAi'(K+\gamma c)C + RBi'(K+\gamma c)D = 0 \quad (56)$$

Next, we write this system of four simultaneous equations as a matrix equation in Eq. 57.

$$\begin{bmatrix} 1 & 0 & -Ai(K) & -Bi(K) \\ 0 & 1 & -RAi'(K) & -RBi'(K) \\ \cos(ka) & \sin(ka) & -Ai(K+\gamma c) & -Bi(K+\gamma c) \\ \sin(ka) & -\cos(ka) & +RAi'(K+\gamma c) & +RBi'(K+\gamma c) \end{bmatrix} \begin{bmatrix} A \\ B \\ C \\ D \end{bmatrix} = \begin{bmatrix} 0 \\ 0 \\ 0 \\ 0 \end{bmatrix} \quad (57)$$

To simplify the notation in this matrix equation, we define the parameters $\Theta$ and X, where $\Theta$ is less than zero with units of radians. Then Eq. 36 is combined with Eq. 59 to obtain Eq. 60 which shows that X is greater than zero. Finally, Eqns. 57, 58, and 59 are combined to obtain Eq. 61, where a is less than zero so $\Theta$ is also less than zero.

$$\Theta \equiv ka \quad (58)$$

$$X \equiv K + \gamma c \quad (59)$$

$$X = \gamma c \frac{E}{V_0} \quad (60)$$

$$\begin{bmatrix} 1 & 0 & -Ai(K) & -Bi(K) \\ 0 & 1 & -RAi'(K) & -RBi'(K) \\ \cos(\Theta) & \sin(\Theta) & -Ai(X) & -Bi(X) \\ \sin(\Theta) & -\cos(\Theta) & +RAi'(X) & +RBi'(X) \end{bmatrix} \begin{bmatrix} A \\ B \\ C \\ D \end{bmatrix} = \begin{bmatrix} 0 \\ 0 \\ 0 \\ 0 \end{bmatrix} \quad (61)$$

Again, note that X is greater than zero, and K is less than zero when E is less than $V_0$ to permit quantum tunneling. The matrix equation defined by Eq. 61 is homogeneous so the determinant of the matrix must be zero, as shown explicitly in Eq. 62 to have a non-trivial solution for the coefficients. Thus, it is shown that the solutions may be determined by varying the parameters R, K, X and $\Theta$ to bring the determinant to zero.

$$Det = \begin{vmatrix} 1 & 0 & -Ai(K) & -Bi(K) \\ 0 & 1 & -RAi'(K) & -RBi'(K) \\ \cos(\Theta) & \sin(\Theta) & -Ai(X) & -Bi(X) \\ \sin(\Theta) & -\cos(\Theta) & +RAi'(X) & +RBi'(X) \end{vmatrix} = 0 \quad (62)$$

There are 24 terms in the determinant, but 10 of them are zero leaving the following 14 in Eq. 63.

$$Det = M_{11}M_{22}M_{33}M_{44} + M_{11}M_{23}M_{34}M_{42} + M_{11}M_{24}M_{32}M_{43} - M_{11}M_{24}M_{33}M_{42}$$
$$- M_{11}M_{23}M_{32}M_{44} - M_{11}M_{22}M_{34}M_{43} + M_{13}M_{24}M_{31}M_{42} + M_{14}M_{22}M_{31}M_{43}$$
$$- M_{14}M_{23}M_{31}M_{42} - M_{13}M_{22}M_{31}M_{44} - M_{13}M_{24}M_{32}M_{41} - M_{14}M_{22}M_{33}M_{41}$$
$$+ M_{14}M_{23}M_{32}M_{41} + M_{13}M_{22}M_{34}M_{41} \quad (63)$$

The terms with indices 11 and 22 in the determinant are both equal to 1 which simplifies the 14 terms as shown in Eq. 64

$$Det = M_{33}M_{44} + M_{23}M_{34}M_{42} + M_{24}M_{32}M_{43} - M_{24}M_{33}M_{42} - M_{23}M_{32}M_{44}$$
$$- M_{34}M_{43} + M_{13}M_{24}M_{31}M_{42} + M_{14}M_{31}M_{43} - M_{14}M_{23}M_{31}M_{42}$$
$$- M_{13}M_{31}M_{44} - M_{13}M_{24}M_{32}M_{41} - M_{14}M_{33}M_{41} + M_{14}M_{23}M_{32}M_{41}$$
$$+ M_{13}M_{34}M_{41} \quad (64)$$

Inserting the expressions for each term from Eq. 62 into Eq. 64 gives the following expression for the determinant as Eq. 65.

$$Det = -RAi(X)Bi'(X) - RAi'(K)Bi(X)\cos(\Theta) - R^2 Ai'(X)Bi'(K)\sin(\Theta)$$
$$+ RAi(X)Bi'(K)\cos(\Theta) + R^2 Ai'(K)Bi'(X)\sin(\Theta) + RAi'(X)Bi(X)$$
$$- RAi(K)Bi'(K)\cos^2(\Theta) - RAi'(X)Bi(K)\cos(\Theta) + RAi'(K)Bi(K)\cos^2(\Theta)$$
$$+ RAi(K)Bi'(X)\cos(\Theta) - RAi(K)Bi'(K)\sin^2(\Theta) - Ai(X)Bi(K)\sin(\Theta)$$
$$+ RAi'(K)Bi(K)\sin^2(\Theta) + Ai(K)Bi(X)\sin(\Theta) \quad (65)$$

Combining the squares of the sine and cosine terms reduces the number of terms from 14 to 12 in Eq. 66.

$$Det = +R^2 Ai'(K)Bi'(X)\sin(\Theta) - R^2 Ai'(X)Bi'(K)\sin(\Theta)$$
$$+ RAi'(X)Bi(X) - RAi(X)Bi'(X)$$
$$+ RAi'(K)Bi(K) - RAi(K)Bi'(K)$$
$$+ RAi(X)Bi'(K)\cos(\Theta) - RAi'(X)Bi(K)\cos(\Theta)$$
$$+ RAi(K)Bi'(X)\cos(\Theta) - RAi'(K)Bi(X)\cos(\Theta)$$
$$+ Ai(K)Bi(X)\sin(\Theta) - Ai(X)Bi(K)\sin(\Theta) \quad (66)$$

Finally, factoring Eq. 66 gives the more functional form in Eq. 67.

$$\begin{aligned}
Det = &+R^2\left[Ai'(K)Bi'(X) - Ai'(X)Bi'(K)\right]\sin(\Theta) \\
&+ R\left[Ai'(X)Bi(X) - Ai(X)Bi'(X)\right] \\
&+ R\left[Ai'(K)Bi(K) - Ai(K)Bi'(K)\right] \\
&+ R\left[Ai(X)Bi'(K) - Ai'(X)Bi(K)\right]\cos(\Theta) \\
&+ R\left[Ai(K)Bi'(X) - Ai'(K)Bi(X)\right]\cos(\Theta) \\
&+ \left[Ai(K)Bi(X) - Ai(X)Bi(K)\right]\sin(\Theta)
\end{aligned} \quad (67)$$

We may determine the null-points for this model by specifying K, X, and R and then varying $\Theta$ to bring the determinant to zero but we may also determine the solution directly from the parameters as follows:

1. Input values for the energy E, the potential $V_0$, and the barrier length c.
2. Use Eq. 68 to determine b, the length for tunneling within the triangular barrier.

$$b = \left(1 - \frac{E}{V_0}\right)c \quad (68)$$

3. Use Eqns. 4, 35, 36 and 52 to determine the corresponding values of k, K, $\gamma$, and R.
4. Form the determinant and iterate using different values of $\Theta$ to bring the determinant to zero.
5. We may only use negative values of $\Theta$ with positive values of k in Eqns. 58 and 4 because $a$, the coordinate at the left end of the model, is a negative quantity.

## V. COEFFICIENTS FOR THE TRIANGULAR BARRIER WAVEFUNCTION.

We have shown that the determinant which is defined in Eq. 62 must be zero. Now we copy the four component equations from Eq. 61 as Eqs. 69, 70, 71 and 72, which are normalized by defining the coefficient A to be unity, and then dividing Eqns. 70 and 72 by $\gamma$.

$$1 - Ai(K - \gamma c)C - Bi(K - \gamma c)D = 0 \quad (69)$$

$$\frac{k}{\gamma}B + Ai'(K - \gamma c)C + Bi'(K - \gamma c)D = 0 \quad (70)$$

$$\cos(ka) + \sin(ka)B - Ai(K - \gamma a)C - Bi(K - \gamma a)D = 0 \quad (71)$$

$$-\frac{k}{\gamma}\sin(ka) + \frac{k}{\gamma}\cos(ka)B + Ai'(K - \gamma a)C + Bi'(K - \gamma a)D = 0 \quad (72)$$

Any three of these four equations may be combined to solve for the normalized coefficients. We combined Eqns. 70 and 71 to obtain Eq. 73, which is used with Eq. 69 to determine C and D, after which Eq. 73 may be used to obtain the third parameter B to complete the solution.

$$\left[\frac{\gamma}{k}Ai'(K-\gamma c)\sin(ka) + Ai(K-\gamma a)\right]C + \left[\frac{\gamma}{k}Bi'(K-\gamma c)\sin(ka) + Bi(K-\gamma a)\right]D = -\cos(ka) \quad (73)$$

## VI. SPECIAL CASE FOR A SHORTED TRIANGULAR BARRIER

Figure 3 shows the special case in which the pre-barrier region is deleted to form a closed circuit having only the shorted triangular barrier. This model may be interpreted as being a tunneling junction shunted by a source of electrical potential which has zero length.

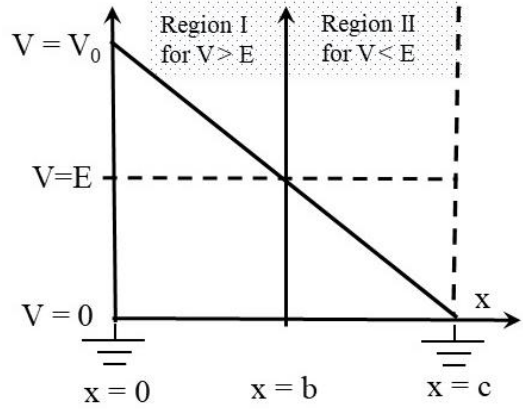

Fig. 3. Potential energy in shorted triangular barrier

Now there is tunneling in Region I and above-barrier transport in Region II. Thus, Θ is zero, so the 12 terms in Eqs. 66 are simplified to give the 8 terms in Eq. 74. These terms are factored to give Eqs. 75 and 76.

$$Det = \begin{bmatrix} Ai'(X)Bi(X) - Ai(X)Bi'(X) + Ai'(K)Bi(K) - Ai(K)Bi'(K) \\ +Ai(X)Bi'(K) - Ai'(X)Bi(K) + Ai(K)Bi'(X) - Ai'(K)Bi(X) \end{bmatrix} R \quad (74)$$

$$Det = \begin{bmatrix} Ai(X)[Bi'(K) - Bi'(X)] + Ai(K)[Bi'(X - Bi'(K))] \\ +Bi(X)[Ai'(X) - Ai'(K)] + Bi(K)[Ai'(K) - Ai'(X)] \end{bmatrix} R \quad (75)$$

$$Det = \begin{bmatrix} [Ai(X) - Ai(K)][Bi'(K) - Bi'(X)] \\ +[Bi(X) - Bi(K)][Ai'(X) - Ai'(K)] \end{bmatrix} R \quad (76)$$

Equation 60 shows that X is greater than or equal to 0, and Eq. 36 shows that K is less than or equal to zero when E is less than $V_0$ to permit quantum tunneling. By inspection of the equations and numerical simulations we have shown that there is only a trivial solution for the special case where X and K are both zero. However, to achieve this limit it would be necessary for c and/or $V_0$ to be zero.

## VII. CLOSED-CIRCUIT MODEL WITH DELTA-FUNCTION BARRIER

This is a second heuristic model but it shows null-point solutions are present in the limit as the barrier width approaches zero as the barrier height becomes infinite.

Figure 3 shows a closed nano-circuit with a delta-function barrier at the origin and a potential of zero elsewhere. As in the previous sections of this paper, the two ground symbols denote a common connection so that "x" represents the coordinate at all of the points in the closed circuit.

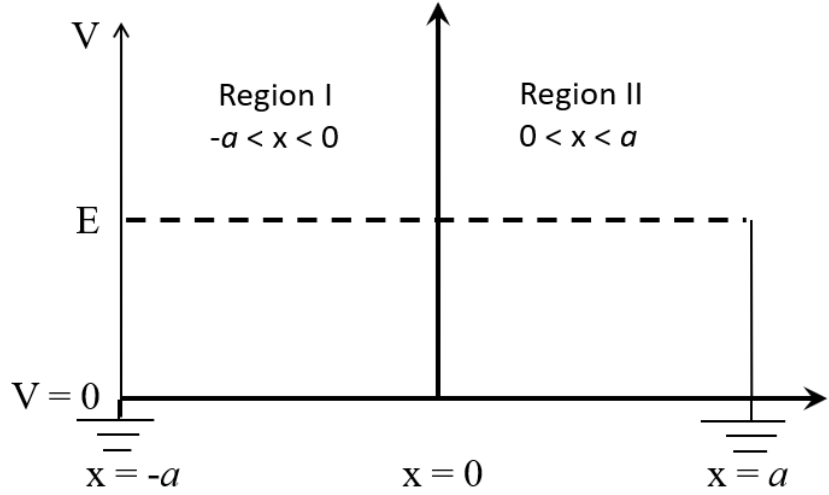

Fig. 4. Closed circuit with a single delta-function barrier.

We use the time-independent Schrödinger equation as Eq. 77 in regions I and II, which is written as Eq. 78 at the origin where α is a constant.

$$-\frac{\hbar^2}{2m}\frac{d^2\psi}{dx^2} + (V-E)\psi = 0 \tag{77}$$

$$-\frac{\hbar^2}{2m}\frac{d^2\psi}{dx^2} - \alpha\delta(x)\psi = E\psi \tag{78}$$

Note that in general, with a delta-function barrier, the coefficients for the wave function are not necessarily the same in the two half=spaces. However, in this case we will see that the coupling between the two ends of the model will be used to simplify the solution. The wavefunctions within each of the two regions are given by Eqs. 79 and 80 where k was defined in Eq. 4. The corresponding derivatives of the wavefunctions in the two regions are given by Eqs. 81 and 82.

$$\psi_I(x) = A\cos(kx) + B\sin(kx) \tag{79}$$
$$\psi_{II}(x) = C\cos(kx) + D\sin(kx) \tag{80}$$
$$\psi_I'(x) = -kA\sin(kx) + kB\cos(kx) \tag{81}$$
$$\psi_{II}'(x) = -kC\sin(kx) + kD\cos(kx) \tag{82}$$

The wavefunction must also be continuous arbitrarily close to the origin so Eqs. 79 and 80 require that C is equal to A, and B is equal to D. We define A equal to unity so Eqs. 79 to 82 are simplified to give Eqs. 83 to 86, leaving only the two unknown coefficients B and D.

$$\psi_I(x) = \cos(kx) + B\sin(kx) \qquad (83)$$

$$\psi_{II}(x) = \cos(kx) + D\sin(kx) \qquad (84)$$

$$\psi_I'(x) = -k\sin(kx) + kB\cos(kx) \qquad (85)$$

$$\psi_{II}'(x) = -k\sin(kx) + kD\cos(kx) \qquad (86)$$

We also require continuity of the wavefunction and its derivative at the two ends of the model because these two points are connected which gives Eqs. 87 to 90 at these two points.

$$\psi_I(-a) = \cos(ka) - B\sin(ka) \qquad (87)$$

$$\psi_{II}(a) = \cos(ka) + D\sin(ka) \qquad (88)$$

$$\psi_I'(-a) = k\sin(ka) + kB\cos(ka) \qquad (89)$$

$$\psi_{II}'(a) = -k\sin(ka) + kD\cos(ka) \qquad (90)$$

Thus, for the wavefunction to be consistent it is necessary to satisfy Eq. 91, which is equivalent to Eq. 92.

$$\cos(ka) - B\sin(ka) = \cos(ka) + D\sin(ka) \qquad (91)$$

$$(B - D)\sin(ka) = 0 \qquad (92)$$

For the derivative of the wavefunction to be consistent it is necessary to satisfy Eq. 93, which is equivalent to Eq. 94.

$$k\sin(ka) + kB\cos(ka) = -k\sin(ka) + kD\cos(ka) \qquad (93)$$

$$2k\sin(ka) + k(B - D)\cos(ka) = 0 \qquad (94)$$

There are only two ways in which Eq. 94 may be satisfied. One is the trivial case that k is zero, which would require that the energy E is zero. The second is that we require that B is equal to D so the sine of $ka$ is zero. Thus, the angle $\Theta$. which is defined as the product $ka$, is an integer multiplied by pi. Thus, the total length of the model, which is $2a$ must be an integer multiple of 2 pi. Thus, to summarize, the boundary conditions are as follows: (1) A is defined as unity. (2) B is zero, (3) C is unity, and D is arbitrary.

## VIII. CLOSED-CIRCUIT MODEL WITH A RECTANGULAR BARRIER VARIED SUCH THAT THE HEIGHT IS INVERSELY PROPORTIONAL TO THE LENGTH.

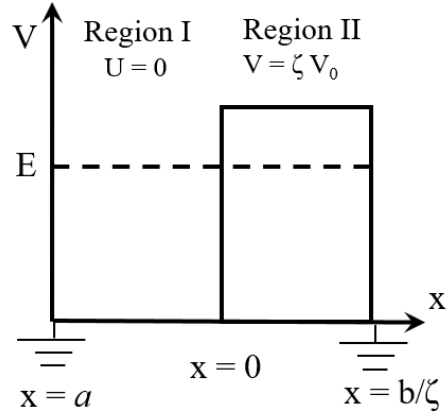

Fig. 5. Potential energy for closed barrier with height varying inversely with length.

Consider the model in Fig. 5, where the barrier height is $V_0$ multiplied by $\xi$ and the barrier length is b divided by $\xi$ so that β in Eq. 5 for Fig. 1 is now β' in Eq. 95.

$$\beta' \equiv \frac{\sqrt{2m(\xi V_0 - E)}}{\hbar} \quad (95)$$

Thus, the determinant in Eq. 15 is changed to that shown as Eq. 96.

$$Det = 4k\beta' + 2(\beta'^2 - k^2)\sinh\left(\beta'\frac{b}{\xi}\right)\sin(ka) - 4k\beta'\cosh\left(\beta'\frac{b}{\xi}\right)\cos(ka) \quad (96)$$

For clarity, we substitute the expression for β' into Eq. 96 to show the full dependence of the determinant on $\xi$ in Eq. 97.

$$Det = 4k\frac{\sqrt{2m(\xi V_0 - E)}}{\hbar} + 2\left[\frac{2m(\xi V_0 - E)}{\hbar^2} - k^2\right]\sinh\left(\sqrt{2m\left(\frac{V_0}{\xi} - \frac{E}{\xi^2}\right)}\frac{b}{\hbar}\right)\sin(ka)$$

$$- 4k\frac{\sqrt{2m(\xi V_0 - E)}}{\hbar}\cosh\left(\sqrt{2m\left(\frac{V_0}{\xi} - \frac{E}{\xi^2}\right)}\frac{b}{\hbar}\right)\cos(ka) \quad (97)$$

As $\xi$ becomes large the arguments of the hyperbolic sine and hyperbolic cosine terms in Eq. 97 approaches zero. Thus, the value of the sinh function will approach its argument and the value of the cosh function will approach unity as shown in Eq. 98.

$$Det = 4k\frac{\sqrt{2m(\xi V_0 - E)}}{\hbar} + \frac{4m(\xi V_0 - E)}{\hbar^2}\sqrt{2m\left(\frac{V_0}{\xi} - \frac{E}{\xi^2}\right)}\frac{b}{\hbar}\sin(ka)$$

$$- 2k^2\sqrt{2m\left(\frac{V_0}{\xi} - \frac{E}{\xi^2}\right)}\frac{b}{\hbar}\sin(ka) - 4k\frac{\sqrt{2m(\xi V_0 - E)}}{\hbar}\cos(ka) \quad (98)$$

Next the E terms are deleted because they are much less than the product of $\xi$ and $V_0$. and the third term is deleted because of its smaller magnitude to obtain Eq. 99.

$$Det = \frac{4k\sqrt{2m\xi V_0}}{\hbar} + \frac{4mV_0 b\sqrt{2m\xi V_0}}{\hbar^3}\sin(ka) - \frac{4k\sqrt{2m\xi V_0}}{\hbar}\cos(ka) \quad (99)$$

Next, the first and third terms of Eq. 99 are regrouped to obtain Eq. 100.

$$Det = \frac{4k\sqrt{2m\xi V_0}}{\hbar}[1 - \cos(ka)] + \frac{4mV_0 b\sqrt{2m\xi V_0}}{\hbar^3}\sin(ka) \quad (100)$$

Next, we replace the product $ka$ with the negative angle $\Theta$, and factor the determinant in Eq. 100 to obtain Eq. 101. Note that under the condition that $\xi$ is large, the determinant is zero when $\Theta$ is an integer multiple of 2 pi radians. Note that this conclusion is necessarily consistent with our earlier derivation with a delta-function barrier in section VI.

$$Det = \frac{4\sqrt{2m\xi V_0}}{\hbar}\left[k[1 - \cos(\Theta)] + \frac{mV_0 b}{\hbar^2}\sin(\Theta)\right] \quad (101)$$

Equation 101 shows that when $\xi$ is large the determinant is zero to provide a solution if and only if Eq. 102 is satisfied. We rewrite as Eq. 102 as Eq. 103 where the LHS has a positive value and the RHS is only a function of the negative angle $\Theta$

$$[1 - \cos(\Theta)] + \frac{mV_0 b}{\hbar^2 k}\sin(\Theta) = 0 \quad (102)$$

$$\frac{mV_0 b}{\hbar^2 k} = \frac{[\cos(\Theta) - 1]}{\sin(\Theta)} \quad (103)$$

Notice that the RHS of Eq. 103 is a periodic function of $\Theta$ with a hyperbolic singularity, becoming large without bound as $\Theta$ approaches minus pi radians from the left, and negatively infinite when approaching this point from the right. Next $\Theta$ may be further changed from minus $\pi$ radians to minus $2\pi$ radians to complete one full cycle of this periodic function. However, the second half of each cycle is non-physical because the right-hand side of Eq. 103 would be negative which is not consistent with the requirement that the left-hand side always has a positive value.

Rearranging Eq. 103 to obtain Eq. 104 shows that, for large values of $\xi$, the product of the potential and the width of the barrier, on the LHS of Eq. 104 is independent of $\xi$ and only depends on $\Theta$. It is explicitly shown in Eq. 105 that this product is the same as that for the unchanged barrier in which $\xi$ is equal to one.

$$(\xi V_0)\left(\frac{b}{\xi}\right) = \hbar^2\sqrt{\frac{2E}{m}}\frac{[\cos(\Theta) - 1]}{\sin(\Theta)} \quad (104)$$

$$V_0 b = \hbar^2\sqrt{\frac{2E}{m}}\frac{[\cos(\Theta) - 1]}{\sin(\Theta)} \quad (105)$$

Again, note that these phenomena repeat as a periodic function over consecutive intervals of $\Theta$ at $(0, -\pi)$, $(-2\pi, -3\pi)$, $(-4\pi, -5\pi)$, etc. The end points for these intervals of $\Theta$ are zero or even multiples of 2 pi radians and there are singularities when $\Theta$ is an odd multiple of 2 pi radians.

## IX. SIMULATIONS FOR THE MODELS

**Table I. Values of the pre-barrier length (minus a, nm), for a square barrier having a potential $V_0$ of 1.0 V, as a function of the particle energy E with five values of the barrier length b in nanometers.**

| E, eV | b = 0.1 nm | b = 0.2 nm | b = 0.5 nm | b = 1.0 nm | b = 2.0 nm |
|---|---|---|---|---|---|
| 0.99 | 0.0579 | 0.116 | 0.289 | 0.528 | 1.157 |
| 0.95 | 0.302 | 0.603 | 1.508 | 3.016 | 6.031 |
| 0.90 | 0.637 | 1.273 | 3.183 | 6.366 | 12.73 |
| 0.85 | 1.01 | 2.02 | 5.06 | 10.11 | 20.22 |
| 0.80 | 1.43 | 2.86 | 7.16 | 14.32 | 28.65 |
| 0.75 | 1.91 | 3.82 | 9.55 | 19.10 | 38.20 |
| 0.70 | 2.46 | 4.91 | 12.28 | 24.56 | 49.11 |
| 0.65 | 3.09 | 6.17 | 15.43 | 30.85 | 61.70 |
| 0.60 | 3.82 | 7.64 | 19.10 | 38.20 | 76.39 |
| 0.55 | 4.69 | 9.38 | 23.44 | 46.88 | 93.76 |
| 0.50 | 5.73 | 11.46 | 28.65 | 57.30 | 114.60 |
| 0.45 | 7.00 | 14.01 | 35.01 | 70.03 | 140.10 |
| 0.40 | 8.59 | 17.19 | 42.97 | 85.94 | 171.90 |
| 0.35 | 10.64 | 21.28 | 53.20 | 106.40 | 212.80 |
| 0.30 | 13.37 | 26.74 | 66.84 | 133.69 | 267.40 |
| 0.25 | 17.19 | 34.38 | 85.94 | 171.89 | 343.80 |
| 0.20 | 22.92 | 45.84 | 114.59 | 229.18 | 458.40 |
| 0.15 | 32.47 | 64.94 | 162.34 | 324.67 | 649.30 |
| 0.10 | 51.57 | 103.13 | 257.83 | 515.65 | 1031.00 |
| 0.05 | 108.62 | 217.72 | 544.30 | 1088.57 | 2177.00 |
| 0.01 | 567.23 | 1134.44 | 2835.00 | 5671.05 | 1135.00 |

**Table II. Values of the pre-barrier length (minus a, nm), for a triangular barrier having a potential $V_0$ of 1.0 V, as a function of the particle energy E with five values of the barrier length c in nanometers.**

| E, eV | c = 0.1 nm | c = 0.2 nm | c = 0.5 nm | c = 1.0 nm | c = 2.0 nm |
|---|---|---|---|---|---|
| 0.99 | 1232.46 | 1232.36 | 1232.06 | 1231.56 | 1230.56 |
| 0.95 | 1258.14 | 1258.04 | 1257.74 | 1257.24 | 1256.24 |
| 0.90 | 1292.62 | 1292.52 | 1292.22 | 1291.72 | 1290.72 |
| 0.85 | 1330.09 | 1329.99 | 1329.69 | 1329.19 | 1328.19 |
| 0.80 | 1371.03 | 1370.93 | 1370.63 | 1370.13 | 1369.13 |
| 0.75 | 1416.00 | 1415.90 | 1415.60 | 1415.10 | 1414.10 |
| 0.70 | 1465.70 | 1465.60 | 1465.30 | 1464.80 | 1463.80 |
| 0.65 | 1521.04 | 1520.94 | 1520.64 | 1520.14 | 1519.14 |
| 0.60 | 1583.15 | 1583.05 | 1582.75 | 1582.25 | 1581.25 |
| 0.55 | 1653.55 | 1653.45 | 1653.15 | 1652.65 | 1651.65 |
| 0.50 | 1734.36 | 1734.36 | 1734.36 | 1734.36 | 1734.36 |
| 0.45 | 1828.17 | 1828.16 | 1828.12 | 1828.07 | 1827.96 |
| 0.40 | 1939.05 | 1939.02 | 1938.95 | 1938.82 | 1938.57 |
| 0.35 | 2072.92 | 2072.87 | 2072.74 | 2072.53 | 2072.10 |
| 0.30 | 2238.98 | 2238.92 | 2238.72 | 2238.38 | 2237.72 |
| 0.25 | 2452.66 | 2452.56 | 2452.26 | 2451.76 | 2450.76 |
| 0.20 | 2742.17 | 2742.07 | 2741.77 | 2741.27 | 2740.27 |
| 0.15 | 3166.40 | 3166.30 | 3166.00 | 3165.50 | 3164.50 |
| 0.10 | 3878.05 | 3877.95 | 3877.65 | 3877.15 | 3876.15 |
| 0.05 | 5484.43 | 5484.33 | 5484.03 | 5483.53 | 5482.53 |
| 0.01 | 12263.69 | 12263.59 | 12263.29 | 12262.79 | 12261.79 |

**Fig 6. Graph of two roots as a function of Θ and particle energy with the triangular barrier b = 0.1 nm, c = 1 nm, and $V_0$ of 1.0 V.**

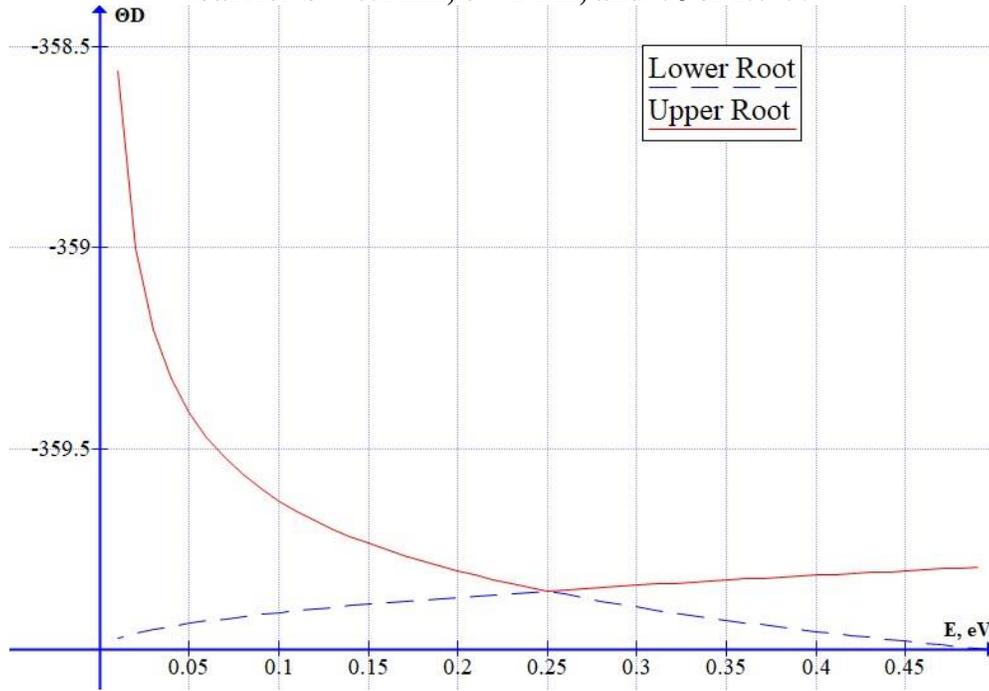

Figure 6 shows the 2 parts for each root that is shown in table II for c = 1.0nm. The vertical displacement between the two curves in figure 6 is extremely small. Thus, we do not see these as two separate lines in the table. A Lease Squares linear regression of the data in figure 6 shows that the angle Θ in degrees has a mean value of -359.77003 with a standard deviation of 0.178557 in degrees.

## X. POSSIBILE APPLICATIONS FOR THE TRIANGULAR BARRIER MODEL

We plan to use a nanowire loop to approximate coherent electron transport in the pre-barrier region with a nano-gap as the tunneling junction. Gall calculated that the longest electron mean-free paths in metals at room temperature are 68.2 nm, 53.3 nm, 39.9 nm, and 37.7 nm, for beryllium, silver, copper, and gold [3], which appear to be suitable. The pre-barrier should be designed to reduce the scattering of electrons at surfaces and grain boundaries. [3],[4] in the pre-barrier region.

While we have only made simulations using the time-independent Schrödinger equation, it is generally possible to make quasistatic approximations for applications when the size of a circuit is much smaller than a wavelength. Thus, we relate the examples to our earlier work at the Los Alamos Laboratory as part of their CINT program. We generated microwave harmonics by focusing a mode-locked laser on the tunneling junction of a scanning tunneling microscope (STM) [5]. These harmonics are at integer multiples of the laser pulse repetition frequency (74.254 MHz) of the laser. A bias-T placed in the tip circuit was used to measure harmonics superimposed on the DC tunneling current. The 200[th] harmonic at 14.85 GHz had a power of only 3.162 x$10^{-18}$ W. However, our analysis suggests that within the STM tunneling junction the harmonics may have extended to terahertz frequencies [6],[7]. The power at these harmonics fell off as the inverse square of their frequencies, but this was caused by capacitive shunting in the measurement circuit. We have described measurement systems that would provide more efficient coupling of the harmonics from the STM and similar methods will be considered in our new application [8].

## XI. SUMMARY AND CONCLUSIONS

In simulations for open-circuit models, such as with a rectangular barrier there is generally an incident wave, a reflected wave, a transmitted wave, and two waves within the barrier. One coefficient, typically that for the incident wave, is specified. Then the system of four independent equations is solved to determine the three remaining coefficients. However, now in each closed-circuit model we have four wave functions and the four equations are dependent because the determinant is zero.

We simulated quantum tunneling in three types of closed circuits. The first one has a rectangular barrier in series with a pre-barrier region. The second has a triangular barrier in series with a pre-barrier region. The third has a delta-function barrier with a pre-barrier region in two connected sections. In each case the four-by-four matrix equation is homogeneous in the four parameters so the determinant must be zero for non-trivial solutions. Thus, with each of the three circuits any three of the four parameters may be specified and the fourth one is varied such that the determinant is zero.

Table I shows the results with the rectangular barrier model where we specify E, $V_0$, and a. The values on each row of the table have two sections for separate types of null-points, where the null-points to the right require much larger values for the pre-barrier length. These two solutions begin at the degenerate case where a is zero and $\Theta$ is either zero or $2\pi$. Since b is a continuous variable, both solutions are on curved lines through the data points.

Table II shows that the null points with a triangular barrier occur when $\Theta$ is equal to $n\pi$ where n is an integer. This is significantly different from the results with the first model using a square barrier. Applications using the triangular barrier would require large pre-barrier lengths or else relatively high energy electrons could be used with higher barriers.

Table III shows that with a delta-function barrier null-points also occur when $\Theta$ is equal to $n\pi$ where n is an integer.

**ACKNOWLEDGMENT**
The author is grateful to our contractor Ezra Pedersen, who used advanced computer methods to extend our original software that is based on this analysis in order to generate the tables and graphs that are in this paper.

**REFERENCES**
1. L. I. Schiff, *Quantum Mechanics*, 3rd ed., New York, McGraw-Hill, New York (1968) 101-104.
2. L.D. Landau and E.M. Lifshitz, *Quantum Mechanics (Non-relativistic Theory)*, 3rd ed., Butterworth-Heinemann, Amsterdam (1977) 101-104.
3. D. Gall, "Electron mean free path in elemental metals", J. Appl. Phys. 119 (2016) 085101.
4. D. Josell, S.H. Brongersma and Z. Tokei, "Size-dependent resistivity in nanoscale interconnects", Annu. Rev. Matter Res. 29 (2009) 231-254.
5. M.J. Hagmann, A.J. Taylor and D.A. Yarotski, "Observation of 200th harmonic with fractional linewidth of $10^{-10}$ in a microwave frequency comb generated in a tunneling junction", Appl. Phys. Lett. 101, 241102 (2012).

6. M.J. Hagmann, F.S. Stenger, and D.A. Yarotski, "Linewidth of the harmonics in a microwave frequency comb generated by focusing a mode-locked ultrafast laser on a tunneling junction", J. Appl. Phys. 114, (2013) 6 pp.

7. M.J. Hagmann, D.G. Coombs and D.A. Yarotski, "Periodically pulsed laser-assisted tunneling may generate terahertz radiation", J. Vac. Sci. Technol. B 35, (2017) 6 pp.

8. M.J. Hagmann and I. Martin, "Design and simulations of a prototype nanocircuit to transmit microwave and terahertz harmonics generated with a mode-locked laser", AIP Advances 12, (2022) 10 pp.